\documentclass[aip,apl,reprint,amsmath,amssymb]{revtex4-1}
\usepackage{graphicx}
\usepackage{dcolumn}
\usepackage{bm}
\usepackage[utf8]{inputenc}
\usepackage[T1]{fontenc}
\usepackage{mathptmx}
\begin{document}
\title{Multimode Parametric Amplifier for K-Band Quantum Measurements}

\author{Chulwoo Ahn}
\email[Corresponding author: ]{chulwoo.ahn@epfl.ch}
\affiliation{Institute of Physics, École Polytechnique Fédérale de Lausanne
(EPFL), CH-1015 Lausanne, Switzerland}
\author{Vladimir Manucharyan}
\affiliation{Institute of Physics, École Polytechnique Fédérale de Lausanne
(EPFL), CH-1015 Lausanne, Switzerland}
\date{October 19, 2026}%

\begin{abstract}
We demonstrate K-band nondegenerate three-wave-mixing amplification
in a multimode resonator comprising an array of 500 SNAILs. Pairing
K-band signal modes with lower-frequency internal idler modes enables
20~dB of gain continuously from 19.6 to 25.6~GHz
with pump frequencies below 32.2~GHz. At 20~dB gain, the 3-dB bandwidth
ranges from 2.6 to 11.7~MHz, and the input-referred 1-dB compression
power reaches $-100.9$~dBm. In the 20--22~GHz band, the average added
noise is 0.73 quanta. A two-mode nondegenerate parametric amplification
model describes the measured gain response and gain--bandwidth behavior,
while a multimode Kerr model captures the overall compression scale.
Our amplifier provides a practical solution for implementing a
state-of-the-art readout of superconducting qubits in the K-band.
\end{abstract}

\maketitle
\raggedbottom
\setcounter{dbltopnumber}{1}
\begin{figure*}[t]
    \centering
    \includegraphics[width=0.98\textwidth]{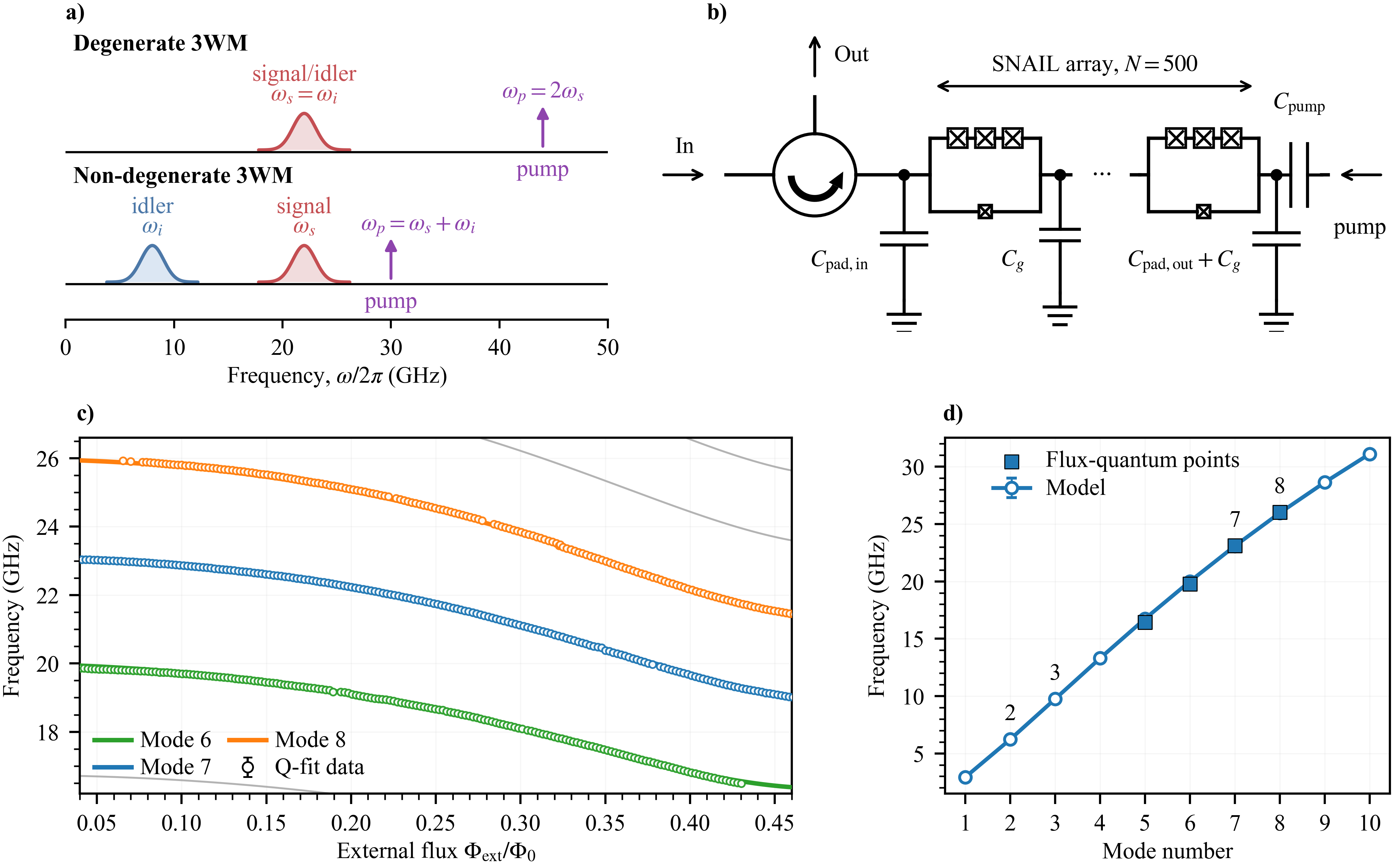}
    \caption{\label{fig:fig1}
    Device concept and multimode spectrum of the distributed SNAIL
    parametric amplifier (SPA). (a) Frequency-domain comparison between
    degenerate three-wave mixing, $\omega_p\simeq2\omega_s$, and the
    nondegenerate process used here,
    $\omega_p\simeq\omega_s+\omega_i$ with $\omega_i<\omega_s$.
    (b) Circuit representation of the short--open nonlinear resonator
    containing 500 SNAIL cells. (c) Flux-dependent spectroscopy of modes
    6--8 (markers) and simultaneous fit to the linear multimode model
    (curves). (d) Calculated zero-flux mode spectrum and independently
    extracted mode-frequency anchors used in the fit.}
\end{figure*}

\begin{figure*}[t]
    \centering
    \includegraphics[width=0.98\textwidth]{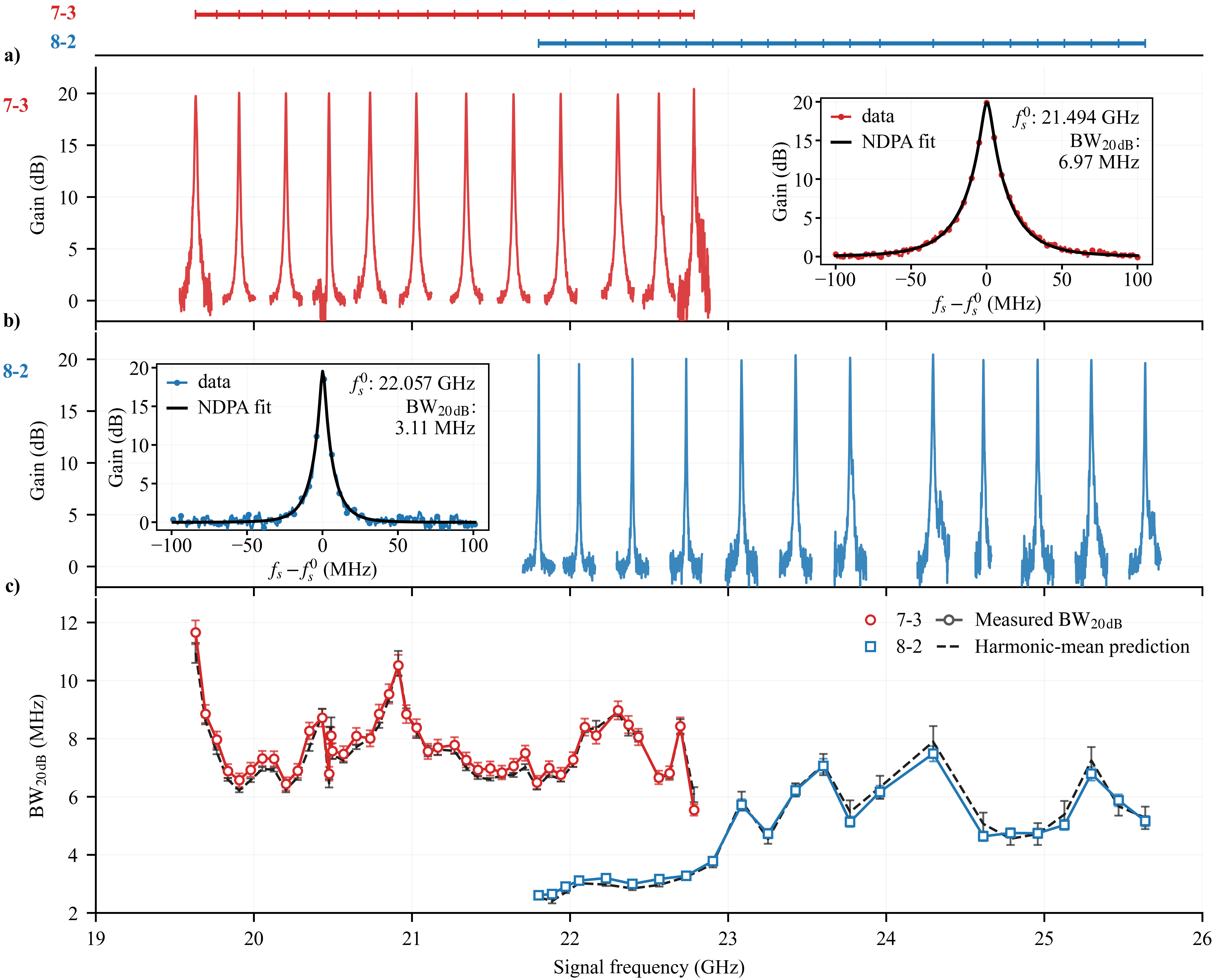}
    \caption{\label{fig:fig2}
    Nondegenerate gain and bandwidth. The upper bars indicate the frequency
    coverage of the 7--3 and 8--2 signal--idler combinations.
    (a),(b) Measured gain traces for the 7--3 and 8--2 combinations;
    each inset compares a measured trace with the corresponding two-mode
    NDPA fit. (c) Measured 3-dB bandwidth normalized to 20-dB peak gain,
    $\mathrm{BW}_{20\,\mathrm{dB}}=\mathrm{BW}\sqrt{G_{\mathrm{peak}}}/10$, and the
    corrected harmonic-mean linewidth calculated from the signal and idler
    linewidths using the full pump-off-normalized model.}
\end{figure*}

Superconducting quantum circuits are increasingly operating at higher
frequencies in pursuit of larger-scale architectures. Raising the
qubit and readout frequencies~\cite{mencia2025raising,anferov2024highfrequency} reduces the equilibrium thermal
occupation at fixed temperature, enabling operation at higher
temperatures where greater cryogenic cooling power is available;
superconducting qubits above 20~GHz have recently been operated above
200~mK~\cite{anferov2024highfrequency}. A readout mode placed far above
the qubit frequency can also suppress drive-induced unwanted state
transitions that limit quantum-nondemolition
fidelity~\cite{nesterov2024mist,kurilovich2025highfrequency}. These
developments require first-stage microwave amplifiers that combine high
gain, low added noise, useful bandwidth, and sufficient dynamic range
in the K band.

Josephson parametric amplifiers (JPAs) provide phase-preserving amplification
near the quantum
limit~\cite{caves1982quantum,roy2016introduction,aumentado2020state}.
In resonator-based JPAs, scaling to higher frequencies by reducing the
inductance or capacitance involves interdependencies among key performance
metrics, including gain, bandwidth, and dynamic range
~\cite{eichler2014dynamic,planat2019saturation}.
Recent experiments have demonstrated resonator-based superconducting
parametric amplification in the K band: Hao \textit{et al.} reported a
waveguide-coupled wireless JPA operating via four-wave mixing and tunable
from 21 to 23.5~GHz~\cite{hao2026wireless}. Junction-free kinetic-inductance
amplifiers have also been demonstrated using NbN resonators near
23~GHz~\cite{wang2026kipa} and at 25~GHz~\cite{zhao2026nbn}, as well as
NbTi complementary split-ring resonators supporting four discrete resonances
spanning 23.3--26.3~GHz~\cite{gilles2026cmwave}. Taken together, these approaches exhibit complementary strengths and motivate architectures that combine broad and continuous frequency tunability, useful bandwidth, practical dynamic range, and near-quantum-limited added noise for K-band operation.

The superconducting nonlinear asymmetric inductive element (SNAIL)
provides a flux-tunable three-wave-mixing
interaction~\cite{frattini2017threewave}. SNAIL-based
three-wave-mixing parametric amplifiers have demonstrated
near-quantum-limited amplification with high dynamic range in the
conventional microwave readout
band~\cite{frattini2018optimizing,sivak2019kerr,
ezenkova2022broadband,dai2025pumpcoupling}.
In four-wave-mixing JPAs, the quartic Kerr nonlinearity enables
the parametric interaction but also produces power-dependent
frequency shifts that can contribute to gain
saturation~\cite{eichler2014dynamic,planat2019saturation}.
Three-wave mixing instead uses the cubic nonlinearity to generate
the desired parametric interaction, allowing the Kerr nonlinearity
to be suppressed and thereby improving dynamic
range~\cite{sivak2019kerr,miano2022frequency}.
Furthermore, the strong pump can be spectrally separated from
the amplified signal~\cite{renbergnilsson2023highgain,dai2025pumpcoupling}.
In degenerate three-wave-mixing amplification, however, amplifying
a signal at frequency $\omega_s$ requires a pump near $2\omega_s$;
operation above 20~GHz therefore requires pump frequencies
above 40~GHz. Generating and delivering such a high-frequency
pump introduces practical challenges associated with transmission
loss, component bandwidth limitations, and parasitic resonances.

A well-studied alternative is nondegenerate three-wave mixing, in
which distinct signal and idler modes are coupled by a sum-frequency
pump~\cite{bergeal2010phase,abdo2013nondegenerate}. In particular, Sivak
\textit{et al.} demonstrated the Josephson array-mode parametric
amplifier (JAMPA), in which different internal modes of a distributed
SNAIL array serve as the signal and idler modes to provide amplification
in the 4--12~GHz band~\cite{sivak2020josephson}. In this Letter, we demonstrate K-band nondegenerate three-wave-mixing
amplification by pairing K-band signal modes with substantially lower-frequency
internal idler modes, so that the sum-frequency pump satisfies
$\omega_p\simeq\omega_s+\omega_i$ and remains well below $2\omega_s$.
Using modes 7 and 8 as signal modes, two overlapping signal--idler
combinations provide approximately 20~dB of gain continuously from 19.6 to
25.6~GHz
while requiring pump frequencies below 32.2~GHz. The device provides
3-dB bandwidths at 20~dB gain from 2.6 to 11.7~MHz and
input-referred 1-dB compression powers from $-111.8$ to
$-100.9$~dBm. For amplification using signal mode 7, the added noise
averaged over 20--22~GHz is $0.73\pm0.15$ quanta, while for signal
mode 8 the corresponding average over 22--25~GHz is $1.83\pm0.91$
quanta, with the values reported as the mean and standard deviation
across the measured frequencies.

Figure~\ref{fig:fig1}(a) contrasts the pump-frequency requirements of
degenerate and nondegenerate three-wave-mixing amplification. We
demonstrate this approach using a distributed short--open resonator
containing 500 SNAIL cells, as shown in Fig.~\ref{fig:fig1}(b). The resonator
supports multiple internal modes that provide the signal and idler
modes required for nondegenerate operation. The device is described by the multimode Hamiltonian
\begin{equation}
\begin{aligned}
\frac{\hat H}{\hbar}
={}& \sum_m\omega_m\hat a_m^\dagger\hat a_m
+\sum_{ijk}g^{(3)}_{ijk}\hat X_i\hat X_j\hat X_k \\
&+\sum_{ijkl}g^{(4)}_{ijkl}\hat X_i\hat X_j\hat X_k\hat X_l
+\cdots ,
\end{aligned}
\label{eq:multimode_hamiltonian_main}
\end{equation}
where $\hat X_m=\hat a_m+\hat a_m^\dagger$ is the dimensionless
quadrature of mode $m$, and $g^{(3)}_{ijk}$ and $g^{(4)}_{ijkl}$
denote the cubic and quartic multimode coupling coefficients,
respectively.

We first determine the underlying linear multimode structure. The
flux-dependent unit-cell inductance is
$L_{\mathrm{cell}}(\Phi)=L_{\mathrm{geo}}+L_{J0}/c_2(\Phi)$, where
$c_2(\Phi)$ is the quadratic coefficient of the SNAIL potential at
its flux-dependent minimum~\cite{frattini2018optimizing}. The reduced
capacitance matrix and boundary
capacitance are fixed by HFSS simulation, and the junction shunt
capacitance density is $45~\mathrm{fF}/\mu\mathrm{m}^2$. In our reduced model, flux
dependence enters the numerical eigenvalue problem only through
$L_{\mathrm{cell}}$~\cite{weissl2015kerr,winkel2020nondegenerate}. A simultaneous fit to the resonance frequencies
extracted from the mode-6, mode-7, and mode-8 spectroscopy data yields
$L_{\mathrm{geo}}=28.9~\mathrm{pH}$,
$L_{J0}=23.5~\mathrm{pH}$, and the SNAIL junction-asymmetry
parameter $\alpha=0.091$. The resulting dispersion and zero-flux spectrum
are shown in Figs.~\ref{fig:fig1}(c) and \ref{fig:fig1}(d).

Guided by the one-tone spectroscopy, we select the 7--3 and 8--2
signal--idler mode pairs, with modes 7 and 8 used as the signal modes.
The gain is defined
as the pump-on reflected power divided by the same-frequency pump-off
power. The 7--3 branch covers 19.6--22.8~GHz with measured peak gains
of 19.8--20.4~dB, while the 8--2 branch covers 21.8--25.6~GHz with
peak gains of 19.6--20.5~dB. The overlapping branches therefore provide
approximately 20~dB of gain without a frequency gap from 19.6 to
25.6~GHz. For each trace, the bandwidth is the full width 3~dB
below its measured peak. Figure~\ref{fig:fig2}(c) reports the corresponding
bandwidth at 20~dB gain. The measured $\mathrm{BW}_{20\,\mathrm{dB}}$ ranges from
5.5 to 11.7~MHz for 7--3 and from 2.6 to 7.5~MHz for 8--2.

The two-mode NDPA model and the bandwidth correction factor used below
are detailed in Supplementary Material, Sec.~I. The signal-mode
linewidth
$\bar{\kappa}_s$ is obtained independently from resonator fits to the one-tone
spectroscopy data, whereas the idler-mode linewidth $\bar{\kappa}_i$ is
determined from the NDPA fit. Representative fitted traces are shown in
the insets of Figure~\ref{fig:fig2}(a) and (b). Although
internal loss can be included as an additional dissipative channel, the
commonly quoted ideal DPA and NDPA gain--bandwidth relations are obtained
in the lossless limit~\cite{eichler2014dynamic,roy2016introduction}.
For our K-band device, internal loss constitutes a non-negligible fraction
of the total signal-mode linewidth. To quantify the resulting
gain--bandwidth behavior, we compare the measured 3-dB bandwidth with
the corrected analytic scale
\begin{equation}
\sqrt{G_{\mathrm{peak}}}\,\mathrm{BW}
\simeq
F_{\mathrm{BW}}
\frac{2\bar{\kappa}_s\bar{\kappa}_i}
{\bar{\kappa}_s+\bar{\kappa}_i},
\label{eq:gain_bandwidth_main}
\end{equation}
where $\bar{\kappa}_m=\kappa_m/(2\pi)$
is the total energy-decay linewidth of mode $m$ in Hz, and
$\mathrm{BW}$ is the full 3-dB bandwidth in Hz.
The unbarred $\kappa_m$ is the energy-decay rate used in the
angular-frequency convention of Supplementary Material, Sec.~I;
external and internal linewidths follow the same $2\pi$ conversion.
The bandwidth correction factor
$F_{\mathrm{BW}}$ accounts for
loss through the signal-mode external coupling fraction
$\alpha_s=\bar{\kappa}_{s,\mathrm{ext}}/\bar{\kappa}_s$, together with pump-off
normalization and finite gain.

\begin{figure}[tb]
\centering
\includegraphics[width=\columnwidth]{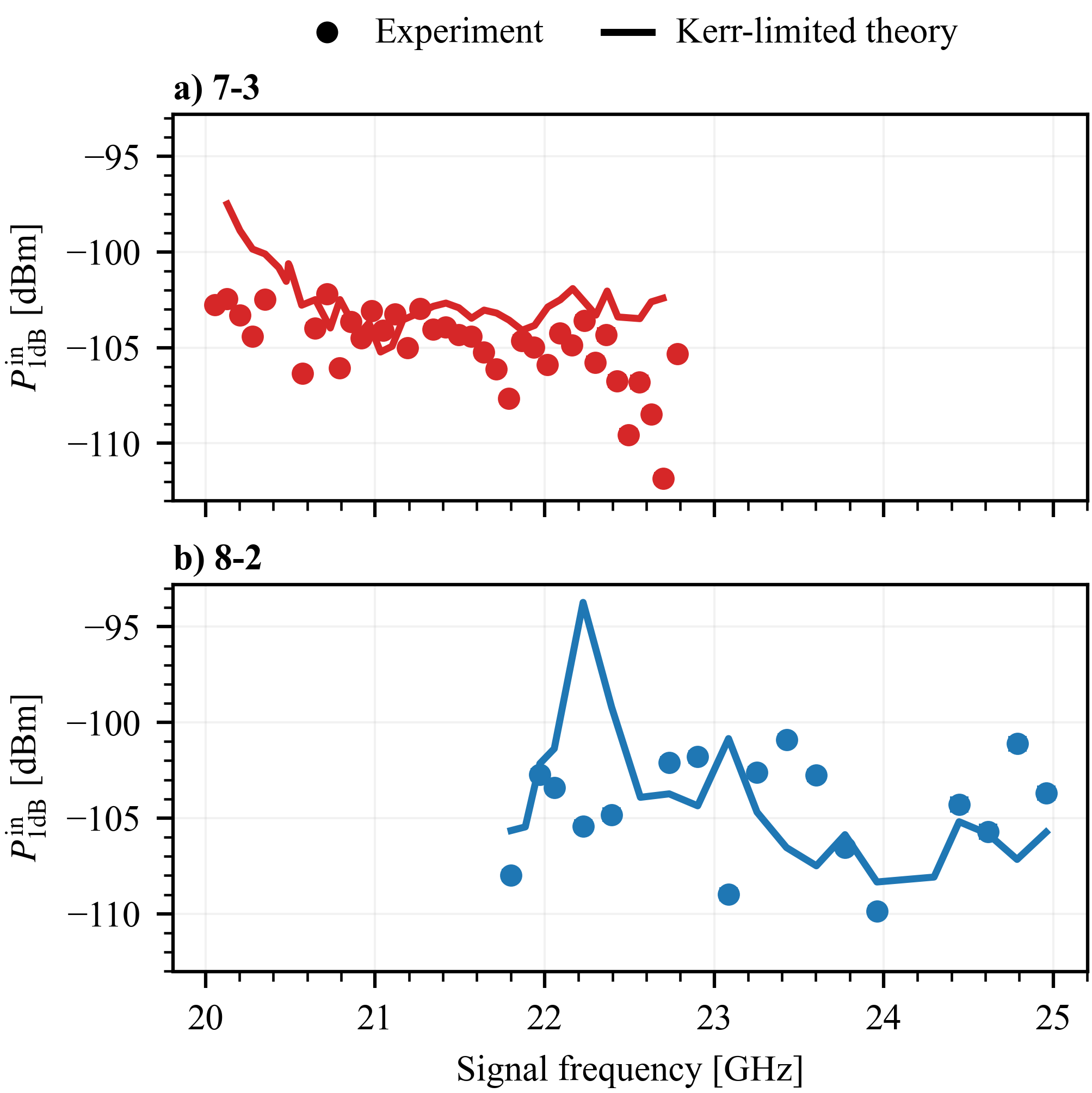}
\caption{\label{fig:fig3}
Input-referred 1-dB compression power versus signal frequency for the
7--3 and 8--2 signal--idler combinations. Markers show measurements
referred to the SPA input plane. Solid curves show the Kerr-limited
calculations including self- and cross-Kerr shifts, with the cubic
correction retaining self-mode and two-mode virtual processes.}
\end{figure}

We next characterize the large-signal response through the
input-referred 1-dB compression power $P_{1\mathrm{dB}}$. The measured
7--3 data span 20.1--22.8~GHz and range from $-111.8$ to
$-102.2$~dBm; the 8--2 data span 21.8--25.0~GHz and range from
$-109.9$ to $-100.9$~dBm. The highest compression power occurs on
the 8--2 branch at a signal frequency of 23.428~GHz and is
$P_{1\mathrm{dB}}=-100.91\pm0.26$~dBm at the SPA input plane, where
the quoted uncertainty includes the propagated calibration terms and
finite-power-step contribution. Details of the input-power calibration
and uncertainty budget are provided in
Supplementary Material, Sec.~II.

To model the measured gain compression, we calculate the multimode Kerr
nonlinearities, which constitute a principal mechanism limiting the dynamic
range of Josephson-junction-based parametric
amplifiers~\cite{eichler2014dynamic,planat2019saturation}. The zero-point
phase-amplitude matrix derived from the mode eigenvectors describes the phase
participation of the modes across the array cells. Together with the
flux-dependent cubic and quartic SNAIL coefficients, this matrix determines
the multimode nonlinear interactions. The finite-power NDPA model includes
the signal and idler self-Kerr coefficients, $K_{ss}$ and $K_{ii}$, and their
cross-Kerr coefficient, $K_{si}$. Each effective Kerr coefficient is written
as $K_{mn}=K_{mn}^{(1)}+K_{mn}^{(2)}$, where $K_{mn}^{(1)}$ is the direct
first-order contribution from the quartic Hamiltonian and $K_{mn}^{(2)}$ is
the second-order contribution generated by virtual transitions through the
cubic Hamiltonian. Following the single-mode treatment of
Ref.~\cite{frattini2018optimizing}, which gives
$K^{(2)}=-60g_3^2/\omega$, we generalize this perturbative calculation to the
multimode spectrum. Our primary calculation retains self-mode $(m,m,m)$ and
two-mode $(m,m,n)$ virtual processes, including their permutations, while
excluding processes involving three distinct modes. The detailed derivation
is given in Supplementary Material, Sec.~II.

We incorporate the calculated Kerr coefficients into the finite-power NDPA
equations through the effective detunings
$\Delta_{s,\mathrm{eff}}=-K_{ss}n_s-K_{si}n_i$ and
$\Delta_{i,\mathrm{eff}}=-K_{ii}n_i-K_{si}n_s$, and solve the steady-state
response as a function of the calibrated input photon flux under the
stiff-pump approximation. The input-referred $P_{1\mathrm{dB}}$ is then
extracted from the calculated gain compression [Fig.~\ref{fig:fig3}]. The
model uses the mode-spectrum eigenvectors, the independently measured signal
linewidth $\bar{\kappa}_s$, and the idler linewidth $\bar{\kappa}_i$ and weak-signal gain
obtained from the NDPA fit. The Kerr-limited model captures the overall compression scale,
while local deviations may reflect pump
depletion~\cite{frattini2018optimizing} or omitted higher-order
nonlinear processes, as well as uncertainties in the model
parameters.

\begin{figure}[tb]
\centering
\includegraphics[width=\columnwidth]{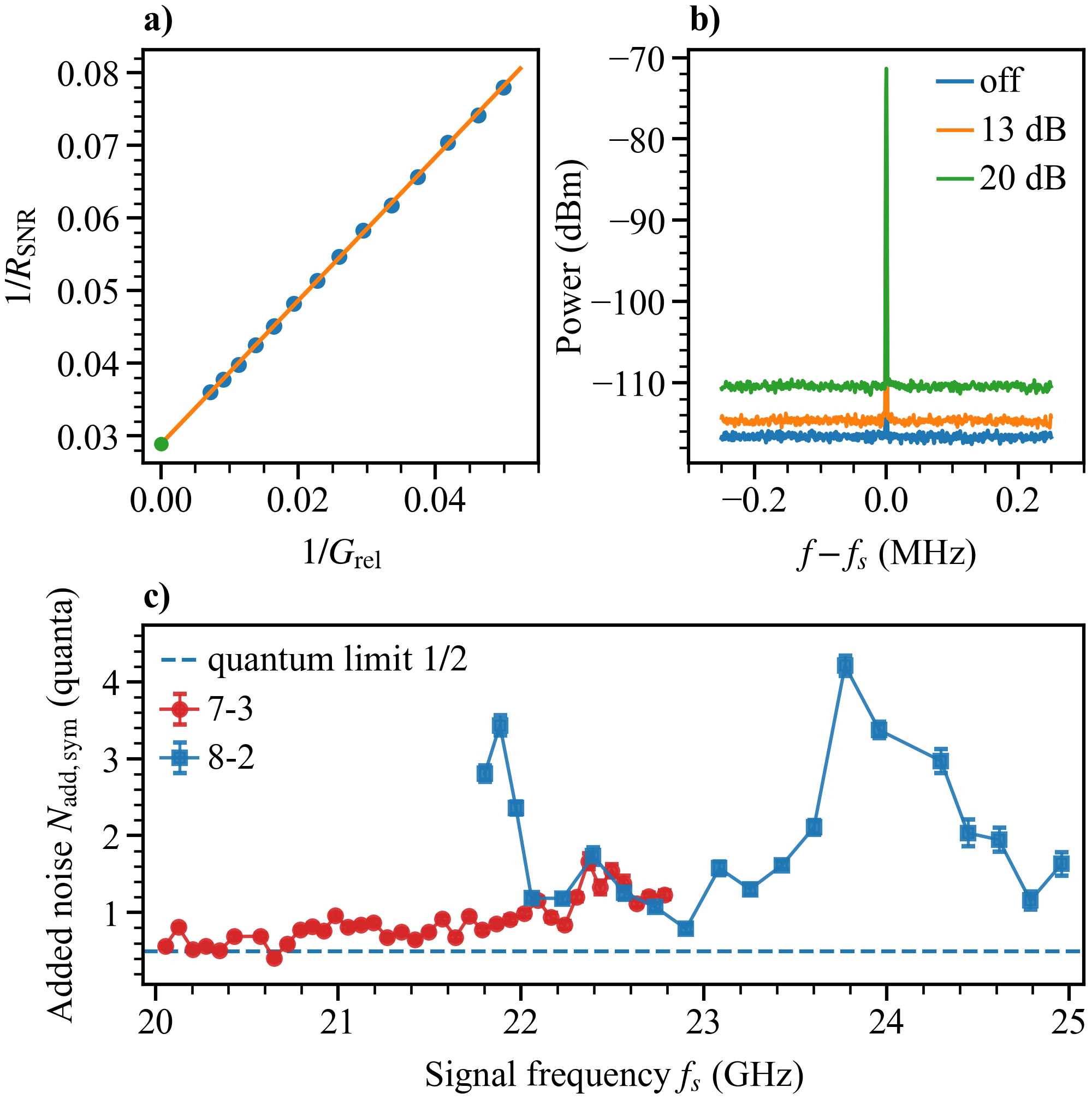}
\caption{\label{fig:fig4}
Input-referred added-noise characterization. (a) Inverse SNR
improvement, $1/R$, versus inverse gain, $1/G$, together with the fitted
finite-gain noise model. (b) Pump-off and representative pump-on
spectrum-analyzer traces for the same operating point.
(c) Input-referred added-noise number versus signal frequency for the
7--3 and 8--2 signal--idler combinations. The dashed line denotes the
phase-preserving quantum limit, $N_{\mathrm{add}}=1/2$. Error bars
propagate uncertainties from the fit, the $N_{\mathrm{rest}}$
calibration, and the pump-off reflection.}
\end{figure}

Finally, we determine the SPA added noise by comparing the gain $G$ with the
SNR improvement, a method widely used for noise calibration of Josephson
parametric
amplifiers~\cite{simoen2015characterization,abdo2013directional}. For
each point, a weak coherent tone and the noise floor are measured while
the pump power is swept. The incident signal noise is written in
symmetrized quanta as
$N_{\mathrm{sig}}=n_{\mathrm{th}}(f_s,T_{\mathrm{input}})+1/2$, where
$n_{\mathrm{th}}=[\exp(hf_s/k_BT_{\mathrm{input}})-1]^{-1}$. The pump-off noise is $N_{\mathrm{off}}=H_{\mathrm{off}}N_{\mathrm{sig}}+
(1-H_{\mathrm{off}})N_{\mathrm{bath}}+N_{\mathrm{rest}}$, where
$H_{\mathrm{off}}=|r_{\mathrm{off}}|^2$ is the pump-off
power-reflection factor, $N_{\mathrm{bath}}$ is the noise associated
with the SPA loss bath, and $N_{\mathrm{rest}}$ is the calibrated
remaining-chain noise referred to the SPA input plane. Assuming that the
loss bath is
equilibrated with the input field,
$N_{\mathrm{bath}}=N_{\mathrm{sig}}$, and therefore
$N_{\mathrm{off}}=N_{\mathrm{sig}}+N_{\mathrm{rest}}$. With the pump
on, the corresponding noise is
$N_{\mathrm{on}}=H_{\mathrm{off}}G N_{\mathrm{sig}}+
(H_{\mathrm{off}}G-1)N_{\mathrm{add}}+N_{\mathrm{rest}}$. The SNR
improvement is
\begin{equation}
R\equiv\frac{\mathrm{SNR}_{\mathrm{on}}}{\mathrm{SNR}_{\mathrm{off}}}
=\frac{G N_{\mathrm{off}}}{N_{\mathrm{on}}}.
\label{eq:snr_improvement_main}
\end{equation}
The inverse SNR improvement is fitted to
$1/R=a/G+b$, as shown in Fig.~\ref{fig:fig4}(a). The added noise is
obtained from the coefficient $b$ as
$N_{\mathrm{add}}=bN_{\mathrm{off}}/H_{\mathrm{off}}-N_{\mathrm{sig}}$.
For a phase-preserving amplifier with a vacuum idler input, the quantum
limit is $N_{\mathrm{add}}=1/2$ quantum~\cite{caves1982quantum,clerk2010quantum}.

The mean added noise across the measured operating points is
$0.73 \pm 0.15$ quanta for the 7--3 branch over 20--22~GHz and
$1.83 \pm 0.91$ quanta for the 8--2 branch over 22--25~GHz, where
the uncertainties denote the standard deviation across frequency
rather than the measurement uncertainty of an individual point.
The absolute values are subject to systematic uncertainties associated
with the frequency-dependent calibration of $N_{\mathrm{rest}}$ and the
insertion-loss correction. The calibration of $N_{\mathrm{rest}}$ is
obtained from temperature-dependent Johnson--Nyquist noise measurements
of a $50~\Omega$ source and referred to the SPA plane using the
frequency-dependent insertion loss between the SPA and the 4-K HEMT.
We therefore
interpret the results conservatively, identifying the 7--3 branch
as approaching the phase-preserving quantum limit within the present
calibration accuracy without interpreting nominal estimates below
the half-quantum bound as evidence of subquantum phase-preserving
amplification.

In conclusion, we have demonstrated a practical K-band parametric
amplifier using an array of 500 SNAILs. The multimode structure of the
SNAIL array enables mixing between K-band signal modes and lower-frequency
internal idler modes, substantially reducing the pump-frequency requirement
while retaining three-wave-mixing operation. Our device combines two
signal--idler mode pairs to provide broad continuous frequency coverage
across most of the K band with approximately 20~dB gain and
near-quantum-limited added noise at selected operating points, addressing
key requirements for first-stage amplification in K-band quantum readout.
Among recently reported resonator-based K-band parametric amplifiers, our
device provides the broadest continuous frequency range at approximately
20~dB gain~\cite{hao2026wireless,wang2026kipa,zhao2026nbn,gilles2026cmwave}.
At the same time, it maintains a balanced combination of bandwidth, power
handling, and added-noise performance. These results establish nondegenerate mixing of widely
separated signal and idler modes as an effective strategy for extending
resonant three-wave-mixing amplification into the K band.

\section*{SUPPLEMENTARY MATERIAL}
See the supplementary material for details of the nondegenerate
gain model and parameter extraction, the Kerr-limited compression
model and input-power calibration, and the added-noise calibration.

\begin{acknowledgments}
We thank Igor Golovchanskiy and Marco Nicolas Mueller for their assistance
with the device fabrication. We thank Han Zheng and Igor Golovchanskiy
for their assistance with wiring the dilution refrigerator. We also thank Raymond
Alonso Mencia, Chien-An Wang, and Taketo Imaizumi for helpful discussions
and feedback on the manuscript. We acknowledge the EPFL Center of
MicroNanoTechnology (CMi) for support with device fabrication. We acknowledge
funding from SEFRI, project number 20QU-1\_215912.
\end{acknowledgments}
\section*{AUTHOR DECLARATIONS}
\subsection*{Conflict of Interest}
The authors have no conflicts to disclose.
\subsection*{Author Contributions}
\textbf{Chulwoo Ahn:} Conceptualization; Methodology; Investigation;
Formal analysis; Software; Data curation; Visualization; Writing -- original draft;
Writing -- review \& editing.

\textbf{Vladimir Manucharyan:} Conceptualization; Funding acquisition;
Project administration; Supervision; Writing -- review \& editing.
\section*{DATA AVAILABILITY}
The data that support the findings of this study are available from the
corresponding author upon reasonable request.
\bibliography{references}
\end{document}


\title{Supplementary Material: Multimode Parametric Amplifier for K-Band Quantum Measurements}

\author{Chulwoo Ahn}
\affiliation{Institute of Physics, École Polytechnique Fédérale de Lausanne
(EPFL), CH-1015 Lausanne, Switzerland}
\author{Vladimir Manucharyan}
\affiliation{Institute of Physics, École Polytechnique Fédérale de Lausanne
(EPFL), CH-1015 Lausanne, Switzerland}
\date{October 19, 2026}%

\maketitle
\renewcommand{\theequation}{S\arabic{equation}}

\section{Nondegenerate gain model and parameter extraction}
\label{app:ndpa_full_derivation}

This section gives the equations used to fit the pump-off-normalized gain
traces, including both external coupling and internal loss. The total linewidth
of mode $m$ is $\kappa_m=\kappa_{m,\mathrm{ext}}+\kappa_{m,\mathrm{int}}$.
Unbarred frequencies, detunings, coupling rates, and linewidths use
the angular-frequency convention unless explicitly divided by $2\pi$. We define
$\bar{\kappa}_m=\kappa_m/(2\pi)$ as the linewidth in Hz, with the
same definition for the external and internal contributions.
Frequencies denoted by $f$ and bandwidths denoted by $\mathrm{BW}$
are expressed in Hz. For a signal probe at frequency $\omega$
and a pump at $\omega_p$, we define
\begin{align}
\Delta_s &= \omega-\omega_{s0},
&
\Delta_i &= \omega_p-\omega-\omega_{i0},
\\
\delta
&=\omega_p-\omega_{s0}-\omega_{i0},
&
\Delta_i&=\delta-\Delta_s .
\label{eq:pub_ndpa_detunings}
\end{align}
Here $\delta$ is the residual sum-frequency detuning. The pump-enhanced
rotating-frame Hamiltonian of the selected signal and
idler modes is
\begin{equation}
\frac{\hat H_{\mathrm{rot}}}{\hbar}
=
-\Delta_s\hat a_s^\dagger\hat a_s
-\Delta_i\hat a_i^\dagger\hat a_i
+\lambda\hat a_s^\dagger\hat a_i^\dagger
+\lambda^*\hat a_s\hat a_i ,
\label{eq:pub_ndpa_hamiltonian}
\end{equation}
where $\lambda$ is the effective three-wave coupling for that pump
state.

\subsection{Linear response and measured gain}

For coherent response, the steady-state Langevin equations are
\begin{subequations}
\label{eq:pub_ndpa_matrix}
\begin{gather}
\begin{pmatrix}
A_s&i\lambda\\
-i\lambda^*&A_i
\end{pmatrix}
\begin{pmatrix}a_s\\a_i^*\end{pmatrix}
=
\begin{pmatrix}
\sqrt{\kappa_{s,\mathrm{ext}}}\,a_{s,\mathrm{in}}\\0
\end{pmatrix},
\label{eq:pub_ndpa_matrix_response}
\\
A_s=\kappa_s/2-i\Delta_s,
\qquad
A_i=\kappa_i/2+i\Delta_i.
\label{eq:pub_ndpa_matrix_definitions}
\end{gather}
\end{subequations}
With $D=A_sA_i-|\lambda|^2$ and
$a_{s,\mathrm{out}}=a_{s,\mathrm{in}}
-\sqrt{\kappa_{s,\mathrm{ext}}}\,a_s$, the pump-on reflection
coefficient is
\begin{equation}
r_{\mathrm{on}}(\omega)
=1-\frac{\kappa_{s,\mathrm{ext}}A_i}{D}.
\label{eq:ndpa_ron}
\end{equation}
The pump-off coefficient follows by setting $\lambda=0$,
\begin{equation}
r_{\mathrm{off}}(\omega)
=
1-\frac{\kappa_{s,\mathrm{ext}}}{\kappa_s/2-i\Delta_s}.
\label{eq:ndpa_roff}
\end{equation}
The experimental gain is the pump-on trace divided by the
pump-off trace at the same frequency,
\begin{equation}
G_{\mathrm{norm}}(\omega)
=
\left|
\frac{r_{\mathrm{on}}(\omega)}{r_{\mathrm{off}}(\omega)}
\right|^2 .
\label{eq:ndpa_normalized_gain}
\end{equation}
All quantitative fits retain the full frequency dependence of
$r_{\mathrm{off}}(\omega)$.

\subsection{NDPA 3-dB bandwidth}

In the narrowband limit, the corrected analytic
gain--bandwidth scale for the pump-off-normalized finite-gain response is
\begin{equation}
\sqrt{G_{\mathrm{peak}}}\,\mathrm{BW}_{\mathrm{3dB}}
\simeq
F_{\mathrm{BW}}
\frac{2\bar{\kappa}_s\bar{\kappa}_i}{\bar{\kappa}_s+\bar{\kappa}_i}.
\label{eq:ndpa_ideal_harmonic_gbw}
\end{equation}
Here the total bandwidth correction factor is
\begin{equation}
F_{\mathrm{BW}}
=
\frac{\alpha_s}{|1-2\alpha_s|}C_{\mathrm{fg}},
\label{eq:ndpa_bandwidth_correction_factor}
\end{equation}
where $\alpha_s=\kappa_{s,\mathrm{ext}}/\kappa_s$ is the signal-mode
external coupling fraction. The finite-gain correction is
$C_{\mathrm{fg}}=A_{\mathrm{abs}}^2/[(A_{\mathrm{abs}}+1)
\sqrt{A_{\mathrm{abs}}^2-2}]$, with
$A_{\mathrm{abs}}=|1-2\alpha_s|\sqrt{G_{\mathrm{peak}}}$.
In the ideal lossless high-gain limit,
$F_{\mathrm{BW}}\rightarrow1$, recovering the usual harmonic-mean
result.
This corrected relation is used only as an analytic reference.
The bandwidth quoted at exactly $20$-dB peak gain is
\begin{equation}
\mathrm{BW}_{20\,\mathrm{dB}}
=
\mathrm{BW}_{\mathrm{3dB}}\frac{\sqrt{G_{\mathrm{peak}}}}{10},
\label{eq:pub_ndpa_bw20}
\end{equation}
where $G_{\mathrm{peak}}$ is linear power gain.

For each trace, $f_s$, $\kappa_s$, and
$\kappa_{s,\mathrm{ext}}$ are determined from the one-tone spectroscopy
resonator fit. The NDPA
fit determines $\kappa_i$, $|\lambda|$, and the residual sum-frequency
detuning $\delta$, which equivalently sets the idler resonance frequency
through $\omega_{i0}=\omega_p-\omega_{s0}-\delta$. Uncertainties from the
one-tone spectroscopy fit are propagated by finite-difference refits and
combined with the local NDPA-fit covariance.

\section{Kerr-limited compression model}
\label{app:ndpa_dynamic_range}

In the main text, the measured 1-dB compression power is interpreted
using a Kerr-limited model that combines direct quartic Kerr terms with
second-order corrections generated by self- and two-mode cubic virtual
processes under the
stiff-pump approximation. The signal linewidth and signal-port external
coupling are fixed by the one-tone spectroscopy resonator fit and carried
into the matched NDPA record, whereas the idler linewidth is determined
from the NDPA fit. The weak-signal gain is taken from the corresponding
low-power compression trace when available, with the NDPA-fit gain used
as a fallback.

\subsection{Kerr-limited stiff-pump model}

For the Kerr-limited curves reported here, we assume
$\Delta_s=\Delta_i=0$ at the weak-signal operating point. The effective
Hamiltonian is
\begin{equation}
\begin{aligned}
\frac{\hat H}{\hbar}
={}&
\lambda_0\hat a_s^\dagger\hat a_i^\dagger
+\lambda_0^*\hat a_s\hat a_i
\\
&+\frac{K_{ss}}{2}\hat a_s^{\dagger2}\hat a_s^2
+\frac{K_{ii}}{2}\hat a_i^{\dagger2}\hat a_i^2
\\
&+K_{si}\hat n_s\hat n_i .
\end{aligned}
\label{eq:pub_dr_kerr_hamiltonian}
\end{equation}
Here $K_{ss}$, $K_{ii}$, and $K_{si}$ denote the signal self-Kerr,
idler self-Kerr, and cross-Kerr coefficients, respectively. For each
operating point, $\lambda_0$ is chosen to reproduce the measured
weak-signal gain in the limit $n_s,n_i\rightarrow0$. For $n_s=|a_s|^2$ and $n_i=|a_i|^2$, the shifts
entering the steady-state equations are
$\delta\omega_s^{K}=K_{ss}n_s+K_{si}n_i$ and
$\delta\omega_i^{K}=K_{ii}n_i+K_{si}n_s$.
The steady-state equations solved by the Kerr-limited model are
\begin{align}
0&=
\left(\frac{\kappa_s}{2}
+i\delta\omega_s^{K}\right)a_s
+i\lambda_0a_i^*
-\sqrt{\kappa_{s,\mathrm{ext}}}\,a_{s,\mathrm{in}},
\\
0&=
\left(\frac{\kappa_i}{2}
-i\delta\omega_i^{K}\right)a_i^*
-i\lambda_0^*a_s ,
\label{eq:pub_dr_steady}
\end{align}
The calibrated device-plane input power sets
$|a_{s,\mathrm{in}}|^2=P_{\mathrm{in}}/(\hbar\omega_s)$.
The four real components of $(a_s,a_i^*)$ are followed by continuation
from low to high input power, selecting the branch connected
continuously to the weak-signal solution.

\subsection{Microscopic nonlinear coefficients}

For array-cell zero-point phases $\theta_{c,m}$, the ordered cubic and
quartic tensors are
\begin{align}
g^{(3)}_{ijk}
&=
\frac{E_J}{\hbar}\frac{\tilde c_3}{3!}
\sum_c\theta_{c,i}\theta_{c,j}\theta_{c,k},
\\
g^{(4)}_{ijkl}
&=
\frac{E_J}{\hbar}\frac{\tilde c_4}{4!}
\sum_c\theta_{c,i}\theta_{c,j}\theta_{c,k}\theta_{c,l}.
\label{eq:pub_dr_tensors}
\end{align}
The ordered tensors enter
$\hat H_3/\hbar=\sum_{ijk}g^{(3)}_{ijk}\hat X_i\hat X_j\hat X_k$,
where $\hat X_m=\hat a_m+\hat a_m^\dagger$. For three distinct modes,
the unique-monomial coefficient is
$g_{sip}\equiv3!\,g^{(3)}_{sip}=6g^{(3)}_{sip}$.
The phase across a linearized array cell is shared between the SNAIL
and its series geometric inductance. Following the nonlinear
current-conservation treatment of Ref.~\cite{frattini2018optimizing},
we define $L_S=L_{J0}/c_2$ and
$p_L=L_S/(L_{\mathrm{geo}}+L_S)$, which give the effective unit-cell
coefficients $\tilde c_3=p_L^3c_3$ and
$\tilde c_4=p_L^4[c_4-3(c_3^2/c_2)(1-p_L)]$.
The first-order perturbative contributions from the quartic Hamiltonian are
$K_{mm}^{(1)}=12g^{(4)}_{mmmm}$ and
$K_{mn}^{(1)}=24g^{(4)}_{mmnn}$.

Off-resonant cubic processes generate second-order dispersive corrections.
For an unperturbed multimode Fock state $|n\rangle$,
\begin{equation}
\delta\omega_{n}^{(2)}
=
\sum_{\mu\ne n}
\frac{|\langle\mu|
\hat H_{3,\mathrm{off}}/\hbar|n\rangle|^2}
{\omega_{n}^{(0)}-\omega_{\mu}^{(0)}} .
\label{eq:pub_dr_second_order}
\end{equation}
Here $\omega_{n}^{(0)}=E_{n}^{(0)}/\hbar$ is the
unperturbed angular frequency associated with the multimode Fock-state
energy $E_{n}^{(0)}$. For a single mode $m$, Eq.~\eqref{eq:pub_dr_second_order}
yields the cubic self-Kerr correction $K_{mm}^{(2)}=-60g_3^2/\omega_m$,
where $g_3\equiv g^{(3)}_{mmm}$~\cite{frattini2018optimizing}.
The operator $\hat H_{3,\mathrm{off}}$ excludes the resonant rotating-wave
term $g_{sip}\hat a_s\hat a_i\hat a_p^\dagger+\mathrm{H.c.}$, which is
already included through the effective three-wave coupling.
The pump mode $p$ is identified as the calculated mode whose frequency is
closest to the measured pump frequency. The perturbative basis contains all
calculated modes up to $N_{\mathrm{max}}=\max(s,i,p)$. Within this basis, we retain the self- and two-mode cubic virtual
processes described in the main text. This restriction is distinct from a
rotating-wave approximation: off-resonant processes involving three distinct
modes can in principle generate second-order dispersive shifts but are omitted
from the reduced model. As a validity criterion for nondegenerate perturbation
theory, a virtual transition is excluded when
$\left|V_{\mu n}/\Delta_{\mu n}\right|>0.1$, where $V_{\mu n}=\langle\mu|\hat H_{3,\mathrm{off}}/\hbar|n\rangle$
and $\Delta_{\mu n}=\omega_n^{(0)}-\omega_\mu^{(0)}$.
This criterion ensures that the perturbative mixing amplitude does not exceed
$0.1$ for any virtual transition included in the second-order sum.

The second-order Kerr contributions follow from finite differences of these
energy shifts,
\begin{align}
K_{ss}^{(2)}
&=\delta\omega_{2s}^{(2)}-2\delta\omega_{1s}^{(2)}
+\delta\omega_0^{(2)},
\\
K_{ii}^{(2)}
&=\delta\omega_{2i}^{(2)}-2\delta\omega_{1i}^{(2)}
+\delta\omega_0^{(2)},
\\
K_{si}^{(2)}
&=\delta\omega_{1s1i}^{(2)}-\delta\omega_{1s}^{(2)}
-\delta\omega_{1i}^{(2)}+\delta\omega_0^{(2)} .
\label{eq:pub_dr_effective_kerr}
\end{align}
The effective Kerr coefficients used in the Kerr-limited model are
$K_{mn}=K_{mn}^{(1)}+K_{mn}^{(2)}$ for $mn\in\{ss,ii,si\}$.

The incident signal power is referred to the SPA input plane using
$P_{\mathrm{SPA,in}}=P_{\mathrm{VNA}}+S_{11,\mathrm{bg}}-G_{\mathrm{rest}}$,
where the powers are expressed in dBm and the correction terms in dB.
Here, $P_{\mathrm{VNA}}$ is the room-temperature source power,
$S_{11,\mathrm{bg}}$ is the resonance-excluded pump-off reflection
background, and $G_{\mathrm{rest}}$ is the power gain of the receiving
chain referred to the SPA output plane. This calibration assumes
unit background power reflection at the SPA,
$|\Gamma_{\mathrm{bg}}|^2=1$; the background is distinct from the
resonant pump-off reflection factor used to normalize the gain.
The reported compression power is estimated as the midpoint of
the interval between the last sampled power below and the first
sampled power reaching 1-dB compression. Treating the calibration
and power-step uncertainties as independent, we calculate
$\sigma_P=(\sigma_{S_{11,\mathrm{bg}}}^2+
\sigma_{G_{\mathrm{rest}}}^2+\sigma_{\mathrm{step}}^2)^{1/2}$,
where $\sigma_{\mathrm{step}}$ is the standard uncertainty associated
with a uniformly distributed crossing position within the sampled
power interval.
This uncertainty does not include common systematic offsets
associated with the unit-background-reflection assumption
or the absolute source-power calibration.

\section{Added-noise calibration}
\label{app:noise}

The added-noise model and the extraction of $N_{\mathrm{add}}$ are
described in the main text. Here we summarize the calibrations of the
remaining-chain noise $N_{\mathrm{rest}}$ and the pump-off
power-reflection factor $H_{\mathrm{off}}$.

\subsection{Remaining-chain calibration}

The remaining measurement chain was calibrated by a Y-factor measurement
using a matched $50~\Omega$ source whose temperature was varied. The
source noise is expressed in photon-normalized units including vacuum
fluctuations,
\begin{equation}
N_T(f,T)
=
\frac{1}{\exp(hf/k_BT)-1}
+
\frac{1}{2}.
\label{eq:pub_noise_NT}
\end{equation}

The measured reference plane is separated from the matched $50~\Omega$
source by the PCB and cryogenic microwave components, whose losses were
measured independently. We define $L_{\mathrm{line}}(f)\geq1$ as the
linear insertion-loss factor between these reference planes. The
remaining-chain noise referred to the $50~\Omega$ source plane is then
\begin{equation}
N_{\mathrm{rest}}^{(50\Omega)}(f)
=L_{\mathrm{line}}(f)N_{\mathrm{rest}}^{(\mathrm{meas})}(f)
+[L_{\mathrm{line}}(f)-1]N_T(f,T_{\mathrm{MXC}}).
\label{eq:pub_noise_Nrest}
\end{equation}
Cable losses were estimated from manufacturer data and extrapolated to
cryogenic temperatures using conductor-loss models spanning the normal
and anomalous skin-effect regimes
\cite{finger2008microwave,reuter1948theory}; a recent comparison for
cryogenic copper structures is given in Ref.~\cite{ratzinger2022anomalous}.
The uncertainty of $N_{\mathrm{rest}}$ includes the Y-factor fit
uncertainty and the uncertainty of this insertion-loss correction.

\subsection{Pump-off reflection}

The pump-off power-reflection factor is determined independently from
pump-off $S_{11}$ spectroscopy. The frequency-dependent background is
estimated after excluding data near the fitted mode trajectories. At a
valid resonator-fit point,
\begin{equation}
H_{\mathrm{off}}
=
\left[
\frac{|S_{11}(f_0,I)|}
{|S_{11,\mathrm{bg}}(f_0)|}
\right]^2 .
\label{eq:noise_Hoff}
\end{equation}
Background standard errors are propagated to $H_{\mathrm{off}}$, and
calibration values are interpolated only within the measured frequency
range.

\bibliography{references}